\documentclass[twocolumn,prl,preprintnumbers,amsmath,amssymb]{revtex4}

\usepackage{graphicx}
\usepackage{times}
\usepackage{amsmath}
\usepackage{amsfonts}
\usepackage{amssymb}
\usepackage{inputenc}
\usepackage{textcomp}

\usepackage{color}
\definecolor{red}{rgb}{1,0,0}
\definecolor{blue}{rgb}{0,0,1}
\definecolor{black}{rgb}{0,0,0}

\begin{document}

\title{The role of momentum-dark excitons in the elementary optical response of bilayer WSe$_{2}$}

\author{Jessica Lindlau$^{1}$, Malte Selig$^{2,3}$, Andre Neumann$^{1}$, L\'{e}o Colombier$^{1}$, Jonghwan Kim$^{4}$, Gunnar Bergh\"auser$^{2}$, Feng Wang$^{4}$, Ermin Malic$^{2}$, and Alexander H\"ogele$^{1}$}

\affiliation{$^1$Fakult\"at f\"ur Physik, Munich Quantum Center,
and Center for NanoScience (CeNS),
Ludwig-Maximilians-Universit\"at M\"unchen,
Geschwister-Scholl-Platz 1, D-80539 M\"unchen, Germany}

\affiliation{$^2$Chalmers University of Technology, Department of
Physics, SE-412 96 Gothenburg, Sweden}

\affiliation{$^3$Institut f\"ur Theoretische Physik, Nichtlineare
Optik und Quantenelektronik, Technische Universit\"at Berlin,
Hardenbergstr. 36, D-10623 Berlin, Germany}

\affiliation{$^4$Department of Physics, University of California
at Berkely, Berkely, 94720 California, U.S.A.}

\date{\today}

\begin{abstract}
Monolayer (ML) transition metal dichalcogenides (TMDs) undergo
substantial changes in the single-particle band structure and
excitonic optical response upon the addition of just one layer. As
opposed to the single-layer limit, the bandgap of bilayer (BL) TMD
semiconductors is indirect which results in reduced
photoluminescence with richly structured spectra that have eluded
a detailed understanding to date. Here, we provide a closed
interpretation of the elementary optical responses of BL WSe$_2$
as a representative material for the wider class of TMD
semiconductors. By combining theoretical calculations with
comprehensive spectroscopy experiments, we identify the crucial
role of momentum-indirect excitons for the understanding of basic
absorption and emission spectra ubiquitously exhibited by various
TMD BLs. Our results shed light on the origin of quantum dot (QD)
formation in monolayer and bilayer crystals and will facilitate
further advances directed at opto-electronic applications of
layered TMD semiconductors in van der Waals heterostructures and
devices.
\end{abstract}

\maketitle

Semiconductor TMDs exhibit remarkable properties in the ML limit,
including a direct bandgap at the K and K$'$ points of the
hexagonal Brillouin-zone (BZ) \cite{Splendiani2010,Mak2010} with
unique spin and valley physics \cite{Xiao2012} of value for novel
opto-valleytronic applications
\cite{Yao2008,Xu2014,Zhang2014,Mak2014,Neumann2017}. In addition
to bright excitons \cite{Chernikov2014,He2014,Ye2014} composed of
electrons and holes in K (or K$'$) valleys with collinear
out-of-plane spin projections, tungsten-based MLs feature
lowest-lying spin-dark excitons
\cite{Zhang2017,Zhou2017,Wang2017,Molas2017} as combinations of K
(or K$'$) electrons and holes of opposite spin
\cite{Liu2013,Kosmider2013a,Kosmider2013b,Kormanyos2014,Kormanyos2015,Echeverry2016}.
The realm of both momentum-direct excitons is expanded by the
notion of momentum-indirect excitons involving electrons and holes
in different valleys
\cite{Wu2015,Dery2015,Qiu2015,Selig2016,Selig2017,Malic2017,Lindlau2017}.
In ML WSe$_2$, for example, momentum-indirect excitons can be
constructed from electrons and holes in opposite K and K$'$
valleys \cite{Dery2015,Selig2017,Lindlau2017}, involve holes in
the $\Gamma$ valley \cite{Malic2017}, or electrons in Q (or Q$'$)
pockets that reside roughly halfway between the $\Gamma$ and K (or
K$'$) points of the first BZ \cite{Selig2017}.

In bilayer TMDs, the single-particle band gap is indirect because
of a downshift of the conduction band (CB) energy at Q well below
K and an upshift of the valence band (VB) edge at the $\Gamma$
point upon the addition of a second layer
\cite{Splendiani2010,Kuc2011,Kumar2012,Cheiwchanchamnangij2012}.
In the specific case of BL WSe$_2$ crystals, the lowest CB minimum
is located at Q, while the VB maximum at K exceeds the one at
$\Gamma$ only by $40 \pm 30$~meV according to angle-resolved
photoemission spectroscopy \cite{Wilson2017}. The associated
photoluminescence (PL) spectra are thus dominated by
momentum-indirect transitions interconnecting electrons and holes
in dissimilar valleys
\cite{Ramasubramaniam2011,Yun2012,Kumar2012,Zhao2013a,Sahin2013,Zhao2013b,Wang2014,Arora2015}.
The BL emission is consistently less efficient, with PL from
short-lived direct excitons \cite{Wang2014} red-shifted by a few
tens of meV from the ML peak emission, and a second peak with
larger red-shift and longer lifetimes \cite{Wang2014} attributed
to momentum-indirect excitons composed of electrons in the K or Q
valleys and holes in the K or $\Gamma$ valleys
\cite{Zhao2013a,Sahin2013,Zhao2013b,Wang2014,Arora2015}. A
detailed understanding of both peaks, however, has remained
elusive \cite{Koperski2017} despite the significance of BL TMDs as
hosts of novel single-photon sources \cite{Kumar2015,Branny2017},
finite valley polarization \cite{Wang2014} or potential
utilization of the spin-layer locking effect in charged BLs
\cite{Jones2014}.

Here we present a comprehensive study, carried out both in
experiment and theory, of exciton manifolds in BL WSe$_2$. Using
cryogenic spectroscopy of BL regions subjected to strain at
unintentional disorder, we identify brightening of
momentum-indirect excitons that in many cases is accompanied by
the formation of quantum dots (QDs) with intense emission of
non-classical light. Complementary experiments reveal the energy
level hierarchy of all excitons involved in determining the
fundamental optical response of BL WSe$_2$. These findings, in
good quantitative agreement with theoretical calculations, not
only explain the intricate details of the BL PL spectra and the
origin of the QD PL, they can be also generalized to other
representatives of TMD materials to facilitate a detailed
understanding of opto-electronic properties of BL and multi-layer
semiconductors.

\begin{figure}[t]
\begin{center}
\includegraphics[scale=1.0]{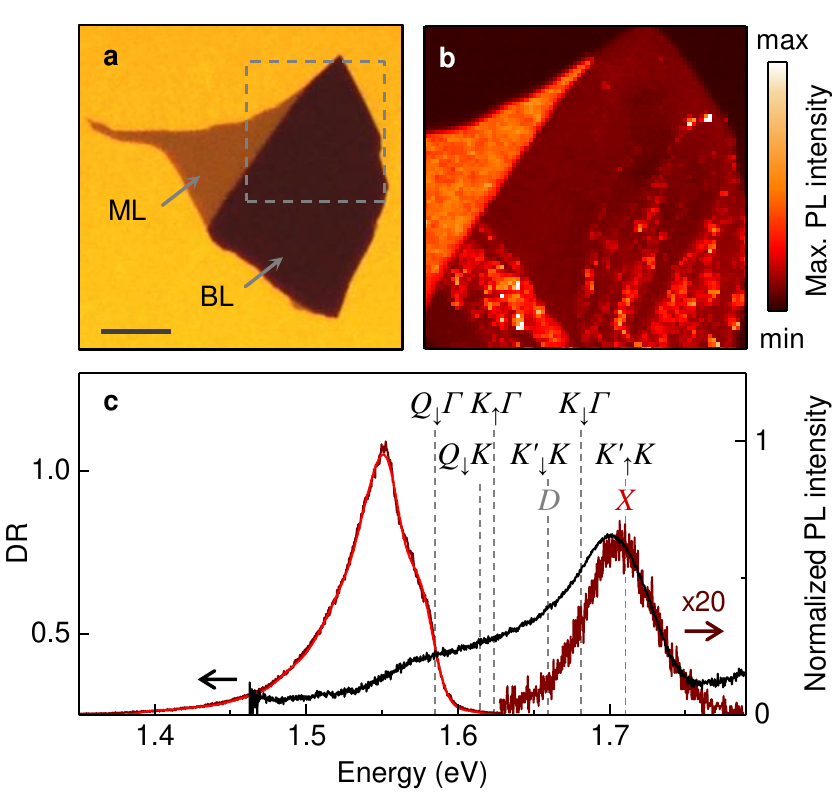}
\caption{\textbf{a}, Optical micrograph of a WSe$_{2}$ flake
exfoliated onto Si/SiO$_2$ with monolayer (ML) and bilayer (BL)
regions indicated by the arrows (the scale bar is $15~\mu$m).
\textbf{b}, Cryogenic raster-scan photoluminescence map of the
upper corner indicated by the dashed square in \textbf{a}.
False-color plot of the photoluminescence maxima in the spectral
range of $1.43 - 1.59$~eV. The bilayer exhibits extended and
punctual regions of brightening attributed to strain at local
folds. \textbf{c}, Differential reflectivity (black) and
normalized photoluminescence spectrum (brown, magnified by a
factor of $20$ in the range of $1.62-1.82$~eV) at a representative
bilayer position away from defects with model fit shown as red
solid line. The energy positions of momentum-bright ($X$ and $D$)
and momentum-dark BL excitons ($Q_{\downarrow}\mathit{\Gamma}$,
$Q_{\downarrow}K$, $K_{\uparrow}\mathit{\Gamma}$,
$K'_{\downarrow}K$, $K_{\downarrow}\mathit{\Gamma}$, and
$K'_{\uparrow}K$, labelled by the capital letters of electron and
hole valleys and the electron out-of-plane spin as subscript) are
indicated by dashed lines. All spectroscopy measurements were
performed at $4.2$~K with excitation at $1.95$~eV.} \label{fig1}
\end{center}
\end{figure}

Cryogenic optical studies (see Supplementary Information for
details) of ML and BL WSe$_2$ were carried out on a flake shown
Fig.~\ref{fig1}a that was obtained by standard exfoliation onto a
Si/SiO$_2$ substrate. Extended ML and BL regions (marked with
arrows) were identified by their respective contrast in the
optical micrograph of Fig.~\ref{fig1}a. The dashed square
indicates the region of the cryogenic hyperspectral raster-scan PL
map recorded with a home-built confocal microscope. The false
color map in Fig.~\ref{fig1}b shows PL peak maxima in the spectral
range of $1.43 - 1.59$~eV, highlighting extended homogeneous
regions of bright and dim ML and BL luminescence, respectively, as
well as distinct BL regions of unintentional disorder with PL
brightening due to local strain \cite{Kumar2015,Branny2017}.

Characteristic PL and differential reflectivity (DR) spectra of BL
WSe$_2$, recorded at $4.2$~K on a representative position away
from disorder, are shown in Fig.~\ref{fig1}c. The PL exhibits a
weak peak around $1.71$~eV and a stronger peak around $1.55$~eV
consistent with previous PL studies of BL WSe$_2$
\cite{Zhao2013a,Zhao2013b,Wang2014,Arora2015}. The peak in DR,
defined as the difference in broad-band reflectivities on and off
the flake normalized to the off-flake reflectivity, confirms the
direct nature of the higher-energy peak in PL. Remarkably, DR
remains at finite values towards lower energies all the way down
to the blue shoulder of the second PL peak. These observations,
combined with theory and comprehensive experiments, allow us to
interpret the optical response of BL WSe$_2$ as originating from
both momentum-direct and momentum-indirect excitons with energy
positions indicated by the dashed lines in Fig.~\ref{fig1}c and
elaborated in the following.

\begin{figure}[t]
\begin{center}
\includegraphics[scale=1.0]{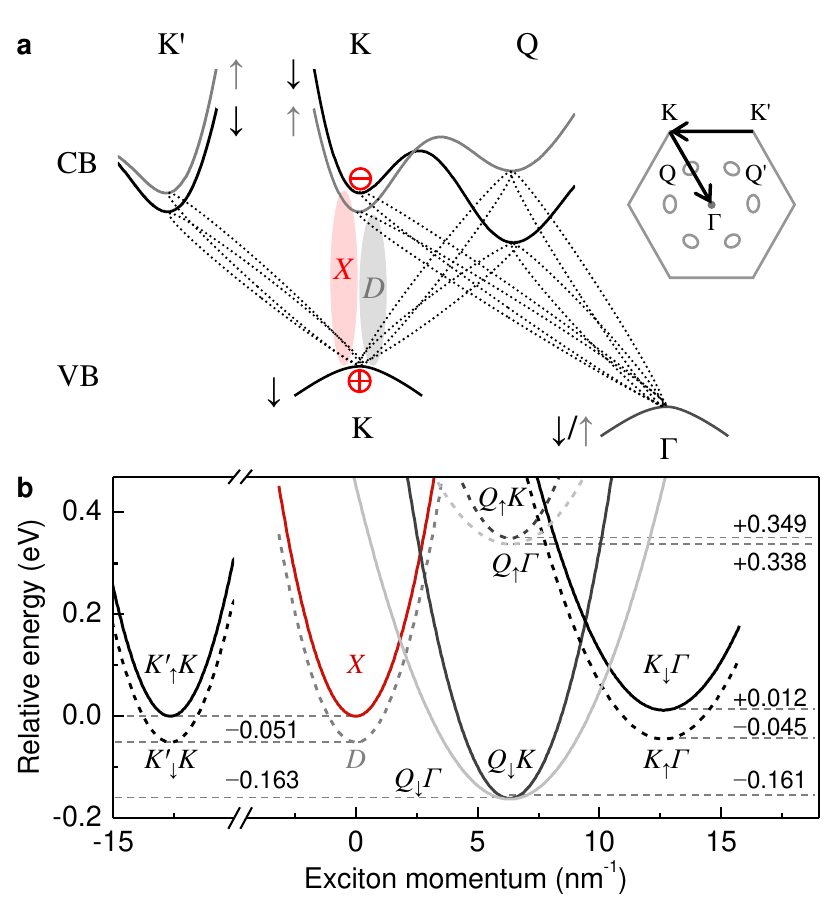}
\caption{ \textbf{a}, Schematic single-particle band diagram of
the conduction and the valence bands of bilayer WSe$_{2}$ along
high symmetry lines of the hexagonal Brillouin zone shown on the
right. Zero-momentum spin-bright ($X$) and spin-dark ($D$)
excitons are formed in the K valley by electrons from spin-down
and spin-up conduction sub-bands indicated in black and grey,
respectively, paired with a spin-down valence band hole.
Momentum-indirect excitons with electrons and holes in dissimilar
valleys are indicated by dashed ellipses. \textbf{b}, Calculated
dispersions of lowest-energy exciton manifolds in BL WSe$_2$ with
energy minima given in eV with respect to the bright exciton $X$.}
\label{fig2}
\end{center}
\end{figure}

To identify all relevant exciton states involved in absorption and
emission, and to interpret the model fit to the lower-energy PL
peak shown as the red solid line in Fig.~\ref{fig1}c, it is
instructive to consider first the single-particle band structure
of BL WSe$_2$ in Fig.~\ref{fig2}a and the associated exciton
dispersions plotted in Fig.~\ref{fig2}b. The relevant states for
the construction of excitons with a hole located at the VB maxima
in K or $\Gamma$ valley (indicated by ellipses in
Fig.~\ref{fig2}a) are the spin-polarized sub-band minima near K, Q
and K$'$ valleys of the CB, with out-of-plane spin projections
indicated by the arrows. We take the spin-degenerate VB maximum at
$\Gamma$ to be $40$~meV below the energy of the spin-polarized
band-edge at K \cite{Wilson2017}, and the energies of the CB at K,
Q and K$'$ from density functional theory calculations
\cite{Wickramaratne2014,Terrones2014}.

The excitonic dispersions, shown in Fig.~\ref{fig2}b, were
computed using the Wannier equation \cite{Kira2006, Haug2004}
within the Keldysh formalism
\cite{Keldysh1978,Cudazzo2011,Berghauser2014}, taking explicitly
the dielectric environment of the TMD material into account. The
corresponding excitons, all of which have their counterparts with
the hole at K$'$, can be separated into the class of zero-momentum
excitons with spin-allowed and spin-forbidden configuration
(labelled as $X$ and $D$, respectively), and finite-momentum
excitons involving Coulomb-correlated electrons and holes from
dissimilar valleys (labelled according to the electron and hole
valleys as capital letters with the electron out-of-plane spin as
subscript Fig.~\ref{fig2}b). All excitons but $X$ are
dipole-forbidden, either due to spin or momentum conservation
constraints.

Energy minima of the branches are given in eV with respect to the
bright exciton $X$ (see Supplementary Information for the details
of theoretical calculations). Consistent with the downshift
(upshift) of the Q ($\Gamma$) valley in the CB (VB) of BL WSe$_2$,
we found the smallest exciton gap for finite-momentum
$Q_{\downarrow}\mathit{\Gamma}$ and $Q_{\downarrow}K$ excitons,
followed by six branches involving an electron in K or K$'$ (two
energy-degenerate branches of $D$ and $K'_{\downarrow}K$ and $X$
and $K'_{\uparrow}K$ excitons with the hole at K, as well as
exciton branches $K_{\downarrow}\mathit{\Gamma}$ and
$K_{\uparrow}\mathit{\Gamma}$ with the hole in at $\Gamma$), and
two branches of excitons with electrons in the spin-up polarized Q
valley $Q_{\uparrow}\mathit{\Gamma}$ and $Q_{\uparrow}K$ with the
hole in ${\Gamma}$ and K, respectively, at highest energies.

Out of these excitons, spin-bright $X$ states emit PL along the
detection axis of our microscope, and the PL from spin-dark $D$
excitons with in-plane emission is detected due to the high
numerical of the objective as well \cite{Wang2017}. In contrast,
all momentum-indirect excitons appear exclusively as phonon
replicas of their optically dark zero phonon line (ZPL) as they
emit photons only with the assistance of acoustic or optical
phonons. With this constraint in mind, we note that the
higher-energy peak of the BL spectrum in Fig.~\ref{fig1}c is
dominated by the ZPL of $X$ (in accord with the onset of strong
DR) with a weak contribution from $D$ to the red wing, while the
lower-energy PL peak is a superposition of phonon sidebands of
momentum-dark excitons $Q_{\downarrow}\mathit{\Gamma}$,
$Q_{\downarrow}K$ and $K_{\uparrow}\mathit{\Gamma}$.

Postponing a detailed explanation for the energy ladder of all
relevant exciton states indicated by the dashed lines in
Fig.~\ref{fig1}c, we first discuss the model fit of the
lower-energy peak in the BL spectrum. For the decomposition of the
peak (red solid line in Fig.~\ref{fig1}c) into the PL
contributions from $Q_{\downarrow}\mathit{\Gamma}$,
$Q_{\downarrow}K$ and $K_{\uparrow}\mathit{\Gamma}$, we set the
energy positions of the respective dark ZPLs to the experimentally
determined values and modelled the phonon replicas by
inhomogeneously broadened Gaussians with a full-width at
half-maximum linewdith $\gamma$. For simplicity, we involved only
one branch of acoustic and optical phonons (the longitudinal
acoustic (LA) and optical (A$_1$) phonon branch) with energies
given in Ref.~\citenum{Jin2014}. Best fit to the spectrum was
obtained with the inhomogeneous linewidth $\gamma=21$~meV. The
inclusion of up to sixth order scattering processes was required
to reproduce the extended low energy tail of the spectrum.

At the level of theory, the energetic ordering of
$Q_{\downarrow}\mathit{\Gamma}$ and $Q_{\downarrow}K$ states is
ambiguous given the small difference of $2$~meV in the energy
minima of the two branches (Fig.~\ref{fig2}b). However,
complementary spectroscopy experiments on strained BL regions and
quantum dots (QDs) discussed in the following remove this
ambiguity and establish the energy scale hierarchy for all
excitons responsible for the elementary optical response of BL
WSe$_2$ with $Q_{\downarrow}\mathit{\Gamma}$ as the lowest-energy
exciton branch, followed by $Q_{\downarrow}K$, $Q_{\downarrow}K$,
$K_{\uparrow}\mathit{\Gamma}$, degenerate $D$ and
$K'_{\downarrow}K$ states, $K_{\downarrow}\mathit{\Gamma}$, and
degenerate $X$ and $K'_{\uparrow}K$ manifolds.

\begin{figure}[t!]
\begin{center}
\includegraphics[scale=1.0]{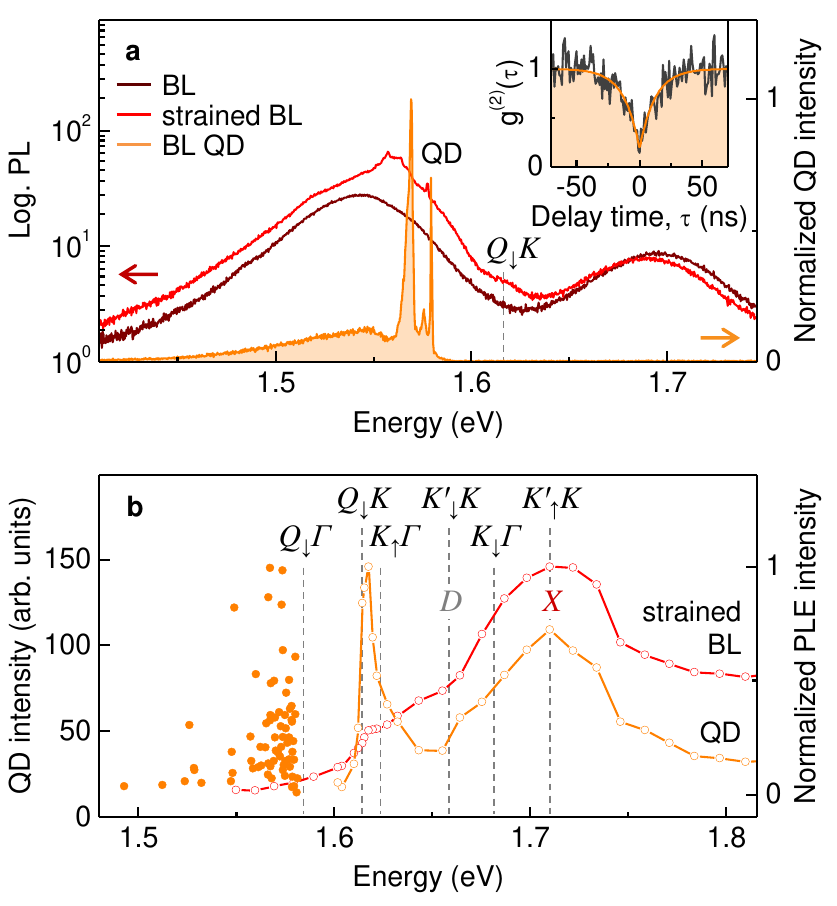}
\caption{\textbf{a}, Photoluminescence from a strained bilayer
region without (red) and with (orange) spectrally narrow and
intense quantum dot (QD) emission recorded at a factor of $1000$
lower excitation power. The bilayer spectrum away from strained
regions (brown) is reproduced from Fig.~\ref{fig1}c for reference.
Note the strain-induced emergence of the shoulder at $1.615$~eV
labelled as $Q_{\downarrow}K$. Inset: typical second-order
coherence of a single quantum dot with pronounced antibunching on
$\sim 10$~ns time scale. \textbf{b}, Distribution of quantum dot
intensities as a function of their peak emission energies (filled
circles, extracted from the map of Fig.~\ref{fig1}b), and
photoluminescence excitation spectra of the quantum dot from
Fig.~\ref{fig3}a and strained bilayer emission (orange and red
open circles, respectively). The dashed lines mark the energy
positions of the relevant exciton states.} \label{fig3}
\end{center}
\end{figure}

The first input to the experimental determination of the exciton
energies is provided by the PL spectroscopy of QDs distributed
randomly along the lines of disorder as in Fig.~\ref{fig1}b. BL
QDs, with intense and spectrally narrow PL emission as in
Fig.~\ref{fig3}a, emerge as a result of local strain
\cite{Kumar2015,Branny2017}. Akin to ML QDs \cite{Srivastava2015a,
He2015, Koperski2015, Chakraborty2015, Kumar2015, Branny2016,
Palacios-Berraquero2017}, the QDs in disordered BL regions were
characterized by strong antibunching signatures in the
second-order correlation function of their PL emission
\cite{Kumar2015,Branny2017}, as demonstrated exemplarily by the
inset data of Fig.~\ref{fig3}a recorded on a different QD with a
dip of $0.2$ at $\tau=0$ and an exponential rise to $1$ on a
timescale of $\sim 10$~ns. By plotting the PL intensity as a
function of the respective energy maximum for all QDs of the
hyperspectral map of Fig.~\ref{fig1}b, we identify a sharp cut-off
to the QD emission energy at $1.584$~eV (indicated by the leftmost dashed
line in Fig.~\ref{fig3}b) which we assign to the state
$Q_{\downarrow}\mathit{\Gamma}$ (see Supplementary Information for
assignment).

The energy position of the next higher-energy momentum-dark state
is revealed by the PL spectroscopy of strained BL regions. The PL
spectrum on a strained position features characteristic blue- and
red-shifts of a few meV for the upper and lower PL peaks (compare
red and brown traces in Fig.~\ref{fig3}a) consistent with $\sim
0.1\%$ of tensile strain which lowers (raises) the CB energy
minimum at K (Q) \cite{Desai2014}. In addition, a shoulder at
$1.615$~eV, indicated by the dashed line in Fig.~\ref{fig3}a, becomes
apparent due to strain-induced brightening of this momentum-dark
transition \cite{Feierabend2017}. The energy position of this
shoulder reappears as a resonance in the photoluminescence
excitation (PLE) spectrum of a strained BL spot (open red circles
in Fig.~\ref{fig3}b). The resonance, marked by the dashed line and
assigned to $Q_{\downarrow}K$, is even more pronounced in the PLE
spectrum of the QD from the same spot position (with the spectrum
in Fig.~\ref{fig3}a) shown by open orange circles in
Fig.~\ref{fig3}b. We note that the PLE spectrum is not QD
specific, it rather represents generic BL resonances in the PLE of
QDs emitting at different observation sites with different
energies (see Supplementary Information for a PLE spectrum of
another QD).

The third successive energy level of momentum-dark states,
identified at $1.624$~eV by the resonance and the shoulder of the
QD and BL PLE spectra of Fig.~\ref{fig3}b, respectively, is
ascribed to $K_{\uparrow}\mathit{\Gamma}$. With this energy, the
experimental values of the three lowest-energy momentum-dark
exciton states can now be hierarchically ordered with respect to
the energy of the bright exciton $X$ at $1.710$~eV deduced from
the peaks of both PLE spectra of Fig.~\ref{fig3}b and from PL and
DR maxima in Fig.~\ref{fig1}c. Referencing all energies to that of
$X$, we note first that the lowest momentum-forbidden state
$Q_{\downarrow}\mathit{\Gamma}$ is red-shifted by $126$~meV
instead of the calculated value of $163$~meV, while the
second-lowest state $Q_{\downarrow}K$ exhibits a red-shift of
$95$~meV instead of $161$~meV expected from theory. Provided that
the effective masses used in the calculations of exciton energies
were correct, these quantitative discrepancies between theory and experiment
convert into an upshift of the CB minimum at the Q valley by
$66$~meV and a downshift of the VB at the $\Gamma$ point by
$29$~meV. Given the uncertainties in band structure calculations
\cite{Terrones2014,Wickramaratne2014} and angle-resolved
photoemission \cite{Wilson2017} used to calculate the exciton
dispersion minima, these corrections of a few tens of meV seem
reasonable.

Finally, with the energies of $X$ and
$K_{\uparrow}\mathit{\Gamma}$ at hand, we estimate the energies of
$D$ and $K_{\downarrow}\mathit{\Gamma}$ in Fig.~\ref{fig3}b by
using the respective spin-orbit splittings of $51$~meV and
$57$~meV from Fig.~\ref{fig2}b. While the energy level of
$K_{\downarrow}\mathit{\Gamma}$ has no compelling signature in the
PLE spectra of Fig.~\ref{fig3}b, the $D$ state coincides with a
clearly pronounced shoulder in the PLE spectrum of the QD. To
complete the energetic ordering of all lowest-lying excitons in BL
WSe$_2$, the states $K'_{\downarrow}K$ and $K'_{\uparrow}K$ are
placed in resonance with $D$ and $X$ by omitting electron-hole
exchange. With this complete set of exciton energies used to
explain the PL signatures in the spectrum of Fig.~\ref{fig1}c, we
can also interpret the onset of DR around $1.550$~eV as stemming
from the first optical sideband of
$Q_{\downarrow}\mathit{\Gamma}$. Absorption due to higher-order
phonon processes are frozen out at the cryogenic temperature of
our experiment, but we expect the absorption onset to become more
and more pronounced towards lower energies with an increasing
population of optical phonons at higher temperatures.

\begin{figure}[t!]
\begin{center}
\includegraphics[scale=1.0]{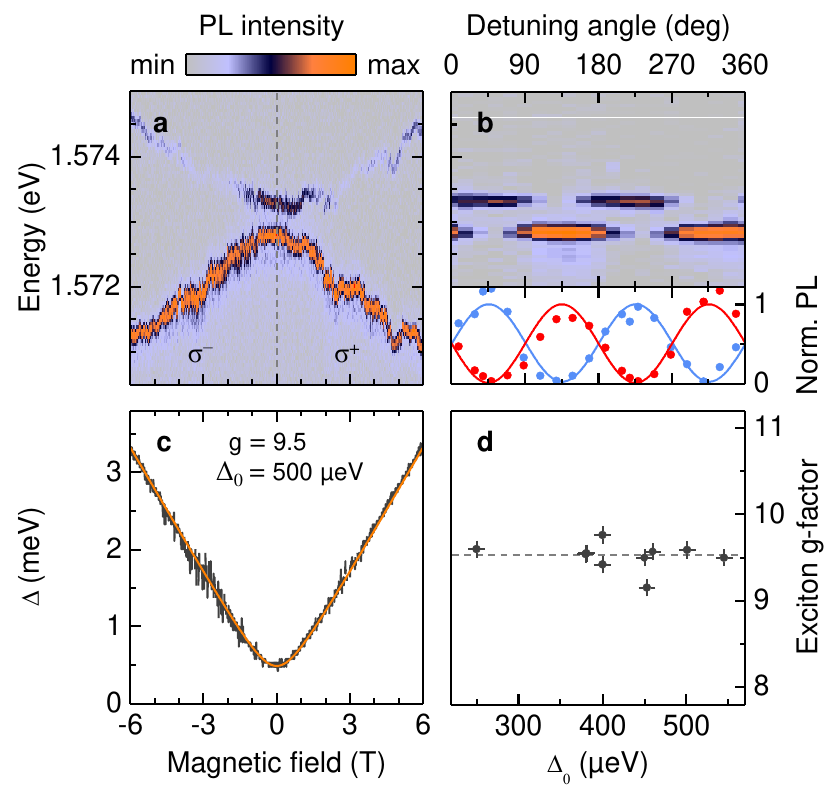}
\caption{\textbf{a}, False-color plot of quantum dot
magnetoluminescence under $\sigma^+$ ($\sigma^-$) polarized excitation for
positive (negative) magnetic fields in Faraday geometry.
\textbf{b}, The quantum dot emission doublet (upper panel) is
characterized by linearly polarized peaks with orthogonal
polarization axes (lower panel; note the anti-correlation in the
intensities of the higher and lower energy peaks shown in red and
blue together with squared sine and cosine fits). \textbf{c},
Energy dispersion of the doublet splitting in magnetic field. Best
fit to the data with a hyperbolic function (solid line) was
obtained for a zero-field fine-structure splitting $\Delta_0$ of
$500$~$\mu$eV and an exciton $g$-factor of $9.5$. \textbf{d},
Distribution of exciton $g$-factors around the mean value of $9.5$
plotted for ten quantum dots with respect to their zero-field
splitting.} \label{fig4}
\end{center}
\end{figure}

This notion of momentum-dark exciton states facilitates a more
insightful perspective on the elementary optical responses of
other bilayer TMD materials and heterobilayer systems
\cite{Forg2017}. Moreover, it provides additional insight into the
origin of QDs in ML \cite{Srivastava2015a, He2015, Koperski2015,
Chakraborty2015, Kumar2015, Branny2016, Palacios-Berraquero2017}
and BL \cite{Kumar2015,Branny2017} TMDs. In addition to spectrally
narrow and bright PL with antibunched photon emission statistics
discussed above, BL QDs share all main signatures of localized
exciton states with ML QDs. In high-resolution micro-PL
spectroscopy, they exhibit a doublet of states with orthogonal
linear polarization (Fig.~\ref{fig4}a and b) which evolves into a
pair of circularly polarized Zeeman-split peaks with increasing
magnetic field (Fig.~\ref{fig4}a and c). The dispersion of the
Zeeman splitting $\Delta$ between the blue and red QD branches
with out-of-plane magnetic field $B$ according to the hyperbolic
function $\Delta=\sqrt{(g\mu_{\text{B}}B)^2+\Delta_0^2}$ (solid
line in Fig.~\ref{fig4}c) is a hallmark of QDs with anisotropic
fine-structure splitting $\Delta_0$ \cite{Bayer2002}. At large
enough fields, the linear asymptote of the Zeeman splitting is
determined by the exciton $g$-factor ($g$) and the Bohr magneton
($\mu_{\text{B}}$).

By applying this analysis to ten randomly selected QDs on strained
BL positions, we extracted $g$ and $\Delta_0$ from the hyperbolic
fit to the Zeeman splitting as for the QD of Fig.~\ref{fig4}a and
c with $g=9.5$ and $\Delta_0=500~\mu$eV. Remarkably, as evidenced
from Fig.~\ref{fig4}d, the $g$-factor of all ten QDs shows only
minor variations around the average value of $9.5 \pm 0.2$
independent of the QD PL energy and despite the spread in the
fine-structure splittings that are typical for ML and BL QDs in
WSe$_2$ in the range of $\sim 400 - 900~\mu$eV
\cite{Srivastava2015a, He2015, Koperski2015,
Chakraborty2015,Kumar2015,Branny2017, Palacios-Berraquero2017}.
This observation suggests that QD excitons are localized
momentum-dark excitons that inherit their $g$-factor from the
delocalized continuum state (i.e. $Q_{\downarrow}\mathit{\Gamma}$
in the case of BL WSe$_2$) and exhibit significant brightening due
to their spread in momentum space upon spatial localization. This
picture is further supported by the sharp cut-off to the emission
energy of BL QDs at the energy of $Q_{\downarrow}\mathit{\Gamma}$
momentum-dark excitons in Fig.~\ref{fig3}b as well as in previous
studies \cite{Kumar2015,Branny2017}.

For QDs in ML WSe$_2$ with similarly sharp cut-off energies at
$\sim 20-25$~meV below the bright state $X$
\cite{Srivastava2015a,Koperski2015,Chakraborty2015,Kumar2015,Branny2017,
Palacios-Berraquero2017} and surprisingly large $g$-factors in the
range of $6-12$ \cite{Srivastava2015a, He2015, Koperski2015,
Chakraborty2015,Kumar2015,Branny2017} this insight suggests the
presence of a momentum-dark reservoir with energy in between the
bright and dark ML excitons $X$ and $D$ as discussed in
Ref.~\citenum{Lindlau2017}. In ML MoSe$_2$ void of momentum-dark
states below the bright exciton, on the other hand, no cut-off
energy to the QD emission was observed and similar values for the
the $g$-factors of QD excitons and the bright exciton $X$ were
found \cite{Branny2016}. To leverage this speculation, theoretical
estimates of exciton $g$-factors would be required for all
excitons constructed from electrons and holes in CB and VB valleys
other than K.


{\bf Acknowledgments:} We thank G.~Cassabois (Universit\'e
Montpellier) and A.~Knorr (TU Berlin) for fruitful discussions.
This research was funded by the European Research Council under
the ERC Grant Agreement No.~336749, the Volkswagen Foundation, and
the German Excellence Initiative via the Nanosystems Initiative
Munich (NIM). A.~H. also acknowledges support from the Center for
NanoScience (CeNS) and LMUinnovativ. E.~M. acknowledges funding
from the European Unions Horizon $2020$ research program under
grant agreement No.~$696656$ - Graphene Flagship as well as from
the Swedish Research Council. M.~S. is thankful to the DFG through
SFB $951$. J.~K. and F.~W. acknowledge support from National
Science Foundation EFRI program (EFMA-1542741).

%





\setcounter{figure}{0} \setcounter{equation}{0}
\renewcommand{\figurename}{Figure~S}
\makeatletter
\def\fnum@figure{\figurename\thefigure}
\makeatother
\newcommand{\figref}[1]{Fig.~S\ref{#1}}

\section{Supplementary Information}


\subsection{Experimental setup}

\vspace{-10pt}

Confocal spectroscopy studies were performed in liquid helium or a
closed-cycle cryostat (attocube systems, attoDRY$1000$) with base
temperatures of $4.2$~K and $3.1$~K, respectively. The latter
cryostat was equipped with a solenoid providing magnetic fields up
to $\pm 9$~T. The sample was positioned with piezo actuators and
scanners (attocube systems, ANP$101$ series and ANSxy$100$/lr)
into the diffraction limited spot of a low-temperature
apochromatic objective with a numerical aperture of $0.82$
(attocube systems, LT-APO/VISIR/$0.82$) and a spot size of
$0.6~\mu$m. A fiber-based home-built microscope, coupled in back
scattering geometry to a standard spectrometer (PI, Acton
SP-$2558$) with a nitrogen-cooled silicon CCD (PI,
Spec-$10$:$100$BR/LN) and a resolution of $0.26$~meV ($0.05$~meV
in Fig.~\ref{fig4} of the main text and in Fig.~S\ref{S1_PLE2},
bottom panel) was used for photoluminescence (PL) and differential
reflectivity (DR) measurements. A supercontinuum laser (NKT
Photonics, SuperK EXW-$12$) was used for DR measurements. The PL
was excited with a continuous wave (cw) diode laser at $639$~nm
(New Focus, Velocity TLB-$6704$), a ps-pulsed diode laser at
$630$~nm (PicoQuant, LDH-P-C-$630$), or a Ti:sapphire laser
(Coherent, Mira $900$) operated either in cw or ps-pulsed mode.
Photoluminescence excitation (PLE) was performed with the
Ti:sapphire laser in cw mode. Two single photon counting avalanche
photo-diodes (PicoQuant, $\tau$-SPAD) were used in a Hanburry
Brown-Twiss configuration for measurements of photon statistics.

\vspace{-10pt}

\subsection{Theoretical calculations}

\vspace{-10pt}

To compute excitonic binding energies for direct as well as for
indirect excitons, we solve the Wannier equation \cite{Kira2006,
Haug2004}
 \begin{equation}
 \frac{\hbar^2 \mathbf{q}^2 }{2m} \varphi_{\mathbf{q}}^{\mu}-\sum_{\mathbf{k}}V^\text{exc}_\mathbf{k-q} \varphi_{\mathbf{k}}^{\mu}+E_\text{gap} \varphi_{\mathbf{q}}^{\mu}=E^{\mu} \varphi_{\mathbf{q}}^{\mu}.
 \end{equation}
Here, $m=m_e m_h/(m_e + m_h)$ denotes the reduced mass with the
electron (hole) mass $m_e$ ($m_h$). The latter were obtained from
density functional theory calculations for electrons
\cite{Wickramaratne2014} and angle-resolved photoemission
spectroscopy measurements for holes \cite{Wilson2017}. The
electronic bandgap $E_\text{gap}$ including band separation for
different valleys and spin bands was obtained from density
functional theory calculations (electrons) \cite{Terrones2014} and
angle-resolved photoemission spectroscopy experiments (holes)
\cite{Wilson2017}. The appearing Coulomb matrix element
$V^\text{exc}_\mathbf{k-q} $ was treated within the Keldysh
formalism
\cite{Keldysh1978,Cudazzo2011,Berghauser2014,Schmidt2016}, where
we can take explicitly into account the dielectric screening from
the environment and the width of the investigated TMD material.
Solving the Wannier equation, we have microscopic access to
exciton eigenenergies $E^{\mu}$ and excitonic wavefunctions
$\varphi^{\mu}_\mathbf{q}$. Note that we do not include exchange
coupling, which is known to lift the degeneracy between spin-like
and spin-unlike exciton states \cite{Qiu2015}.

\vspace{-10pt}

\subsection{Discussion of PLE spectra and peak ordering}

\vspace{-10pt}

\begin{figure}[t]
\begin{center}
\includegraphics[scale=1.0]{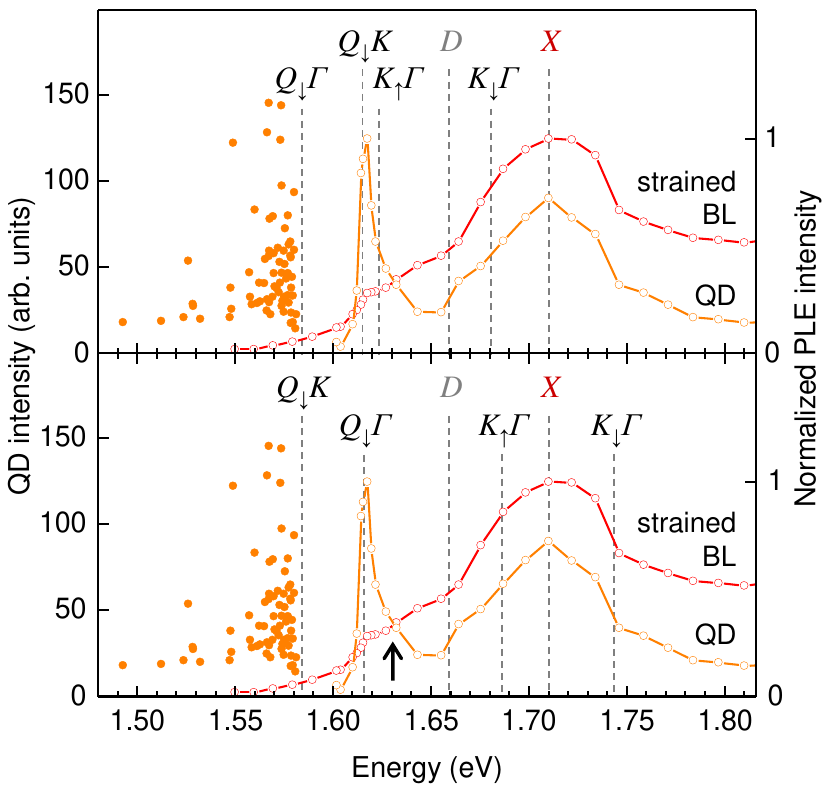}
\caption{Data duplicated from Fig.~3b of the main text with
exciton labelling same as in the main text. Upper panel: energy
positions and peak assignments of dark-exciton states according to
the theoretically predicted energy scale hierarchy. Lower panel:
energy positions of momentum-dark excitons according to the
reversed ordering of the two lowest-energy states. Note the
missing peak assignment in the lower panel indicated by the
arrow.} \label{S1_PLE1}
\end{center}
\end{figure}

\begin{figure}[t!]
\begin{center}
\includegraphics[scale=1.0]{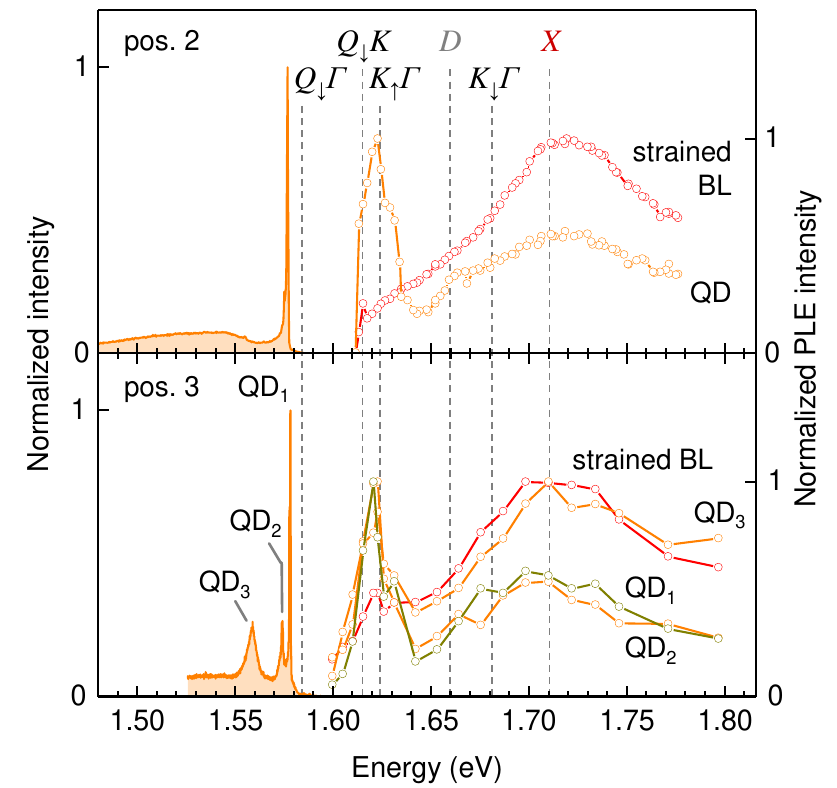}
\caption{Same as in Fig.~S\ref{S1_PLE1} but recorded on a
different position of disorder-strained bilayer WSe$_2$ with
quantum dot emission. The exciton energies marked by the dashed
lines are the same as in the upper panel of Fig.~S\ref{S1_PLE1}
and in Fig.~3b of the main text.} \label{S1_PLE2}
\end{center}
\end{figure}

In order to rationalize the energetic ordering of momentum-dark
excitons, the data Fig.~3b of the main text is reproduced in both
panels of Fig.~S\ref{S1_PLE1}. The energies identified from the
cut-off to the quantum dot (QD) PL as well as from QD and bilayer
(BL) PLE spectra were referenced to the energy of the bright
exciton $X$ at $1.710$~eV that lies $51$~meV above its spin-orbit
split momentum-direct spin-dark counterpart $D$ according to
theory. The energies of the two lowest-energy dark-exciton states
were identified as $E_1=1.584$~eV and $E_2=1.615$~eV,
respectively, yielding experimental shifts of
$\Delta^\text{exp}_{1-X}=E_1-E_X=-126$~meV and
$\Delta^\text{exp}_{2-X}=E_2-E_X=-95$~meV with respect to the
energy of $X$. According to theoretical calculations, the two
lowest-energy states $Q_{\downarrow}\mathit{\Gamma}$ and
$Q_{\downarrow}K$, separated by
$\Delta^\text{th}_{Q\Gamma-X}=-163$~meV and
$\Delta^\text{th}_{QK-X}=-161$~meV, respectively, compete for the
assignment to the lowest-energy states.

First, we test the scenario of preserved energy scale hierarchy
with $Q_{\downarrow}\mathit{\Gamma}$ state being lowest in energy
(with energy $E_1$), followed by the state $Q_{\downarrow}K$ (with
energy $E_2$). We note that the states $Q_{\downarrow}K$ and $X$
share their hole in the K valley and thus the energy difference
can be entirely attributed to the electron in the CB minimum at Q.
With $\Delta^\text{exp}_{2-X}-\Delta^\text{th}_{QK-X}=66$~meV, we thus directly
obtain the upshift of the exciton energy with the hole in K and
the electron in Q. For the state $Q_{\downarrow}\mathit{\Gamma}$
we obtain the difference between the experimental and theoretical
values as $\Delta^\text{exp}_{1-X}-\Delta^\text{th}_{Q\Gamma-X}=37$~meV, which
implies a downshift of the VB maximum at $\Gamma$ by $29$~meV by
using the upshift of the CB minimum at Q calculated above. The
energy of the state $K_{\downarrow}\mathit{\Gamma}$, which shares
with $X$ the electron in the spin-up polarized sub-band at K,
computes by including the upshift of $\Gamma$ to $1.681$~eV.
Finally, the energy of the state $K_{\uparrow}\mathit{\Gamma}$ is
obtained as $1.624$~eV by taking into account the theoretically
calculated spin-orbit splitting of $57$~meV between
$K_{\downarrow}\mathit{\Gamma}$ and $K_{\uparrow}\mathit{\Gamma}$.

The second scenario probes the reversed ordering where the state
$Q_{\downarrow}K$ is lowest (with energy $E_1$) and
$Q_{\downarrow}\mathit{\Gamma}$ is second-lowest state (with
energy $E_2$). Calculations of the respective energies for all
relevant momentum-dark states along the lines of arguments given
above yields an upshift of the Q valley by $35$~meV and an upshift
of the $\Gamma$ valley by $33$~meV. Accordingly, the energies of
$K_{\downarrow}\mathit{\Gamma}$ and $K_{\uparrow}\mathit{\Gamma}$
states are obtained as $1.743$~eV and $1.686$~eV, respectively.

The energy positions for all relevant excitons obtained from the
two competing assignment scenarios are plotted as dashed lines in
Fig.~S\ref{S1_PLE1}. The upper panel shows the energetic ordering
in accord with preserved hierarchy, while the lower panel shows
the results of reversed ordering. The failure of the latter to
predict the resonance in PLE at $1.624$~eV (indicated by the black
arrow in the lower panel) which is consistently ascribed in the
framework of the former to the state
$K_{\uparrow}\mathit{\Gamma}$, provides strong evidence for
$Q_{\downarrow}\mathit{\Gamma}$ being the lowest and
$Q_{\downarrow}\mathit{K}$ being the second-lowest state.

Remarkably, all PLE resonances appear at the same energy positions
for a different BL region subjected to unintentional strain (red
PLE spectrum in Fig.~S\ref{S1_PLE2}) with confocal PL from a QD
emitting at a different PL energy (orange PLE spectrum in
Fig.~S\ref{S1_PLE2}). This observation indicates that the PLE
resonances are not QD-specific (e.g. due to excited QD states that
would differ from dot to dot because of different confinement
potentials) but indeed probe the absorption of BL WSe$_2$.

\end{document}